# A COMPARISON OF NEW CALCULATIONS OF $^{10}$Be PRODUCTION IN THE EARTHS POLAR ATMOSPHERE BY COSMIC RAYS WITH $^{10}$Be CONCENTRATION MEASUREMENTS IN POLAR ICE CORES BETWEEN 1939-2005 – A TROUBLING LACK OF CONCORDANCE PAPER #1

W.R. Webber[1] and P.R. Higbie[2]

1. New Mexico State University, Department of Astronomy, Las Cruces, NM 88003, USA
2. New Mexico State University, Physics Department, Las Cruces, NM 88003, USA

**ABSTRACT**

19

20     Using new calculations of [10]Be production in the Earths atmosphere which are based on

21     direct measurements of the 11-year solar modulation effects on galactic cosmic rays and

22     spacecraft measurements of the cosmic ray energy spectrum, we have calculated the yearly

23     average production of [10]Be in the Earths atmosphere by galactic and solar cosmic rays since

24     1939.  During the last six 11-year cycles the average amplitude of these production changes is

25     36%.  These predictions are compared with measurements of [10]Be concentration in polar ice

26     cores in both the Northern and Southern hemisphere over the same time period.  We find a large

27     scatter between the predicted and measured yearly average data sets and a low cross correlation

28     ~0.30.  Also the normalized regression line slope between [10]Be production changes and [10]Be

29     concentration changes is found to be only 0.4-0.6; much less than the value of 1.0 expected for a

30     simple proportionality between these quantities, as is typically used for historical projections of

31     the relationship between [10]Be concentration and solar activity.  The distribution of yearly

32     averages in the [10]Be concentration level in the data from the Dye-3 ice core in Greenland for the

33     time period 1939-1985, contains a "spike" of high concentration one year averages which is not

34     seen in the production calculations.  These and other features suggest that galactic cosmic ray

35     intensity changes which affect the production of [10]Be in the Earths atmosphere are not the sole

36     source of the [10]Be concentration changes and confirm the importance of other effects, for

37     example local and regional climatic effects, which could be of the same magnitude as the [10]Be

38     production changes.

39

## **Introduction**

The study of the $^{10}$Be concentration in polar ice cores has become an important and widely used new tool to probe the history of solar activity and its interaction with the Earth through the solar modulation of cosmic rays (e.g., Beer, et al., 1990, 1998). This solar activity is typically reconstructed by assuming a simple linear relationship between $^{10}$Be concentration in ice cores and the rate of $^{10}$Be production in the atmosphere (e.g., Bard, et al, 2000; McCracken and McDonald, 2001; Usoskin, et al., 2003; McCracken, et al., 2004; Vonmoos, et al., 2006; Muschaeler, et al., 2007). In the reconstructions of these historical $^{10}$Be nuclei production rates, various solar activity indices, such as the sunspot number have typically been used as proxies for determining the actual $^{10}$Be production rates. In this paper we have used newly calculated atmospheric production rates of $^{10}$Be by both galactic and solar cosmic rays (Webber and Higbie, 2003; Webber, Higbie and McCracken, 2007) to examine directly for the first time the correlation between $^{10}$Be production and the measured $^{10}$Be concentration in ice cores over the time period from 1939 to the present.

The studies using cosmogenic nuclei such as $^{10}$Be as well as $^{14}$C, $^{3}$H, $^{36}$Cl and $^{7}$Be, have been made possible by the advent of AMS which permits the detection of very small samples of trace isotopes produced by cosmic rays incident on the top of the Earths atmosphere to a statistical accuracy ~5-10% per measurement period. $^{10}$Be is particularly valuable in this sense because it is believe to precipitate out of the atmosphere in ~1 year or less, mostly in the form of rain or snow whereas $^{14}$C, for example, has a much longer and more involved sequestering process. $^{10}$Be as well as other cosmogenic isotopes are produced by galactic cosmic ray (proton and heavier nuclei) interactions with the atmospheric elements $^{14}$N, $^{16}$O and $^{40}$Ar. Based on direct studies of cosmic ray time variations during the last 60-70 years using balloons and spacecraft, covering six 11-year solar activity cycles, this production undergoes certain well defined changes. In the N-S Polar Regions, above ~65 degrees geomagnetic latitude, there is an average 11-year solar modulation of the production of all of the cosmogenic isotopes including $^{10}$Be by ~36%. The global average of this production exhibits a weaker, but still significant 11-year modulation (Webber, Higbie and McCracken, 2007). Thus depending on the level of atmospheric mixing and large scale climatic effects one might measure $^{10}$Be concentrations in ice cores in the polar regions that reflect the full 11-year modulation of ~36% or a weaker global average concentration with a smaller 11-year modulation.

71    In this paper we compare calculations of the atmospheric cosmic ray production of [10]Be

72    nuclei over the last ~70 years, based on direct measurements of cosmic ray intensities at the top

73    of the Earth's atmosphere (Webber and Higbie, 2003; Webber, Higbie and McCracken, 2007),

74    with corresponding high latitude [10]Be concentration measurements made over the same time

75    period.

76    The [10]Be concentration measurements include ice core measurements in both the N and S

77    Polar Regions.  In the northern hemisphere there is the Dye 3 measurement at ~65° latitude in

78    Greenland (Beer, et al., 1990).  This measurement has a basic time resolution of 1 year and

79    covers an extended time period up to 1985.  This same location also has ice pit data on a finer

80    time scale from 1977 to 1984 (Beer, et al., 1991).  We should also mention the GRIP ice core

81    data from Summit Greenland at a latitude of ~78° (Yiou, et al., 1997).  This data has a coarser

82    time scale and is therefore not as useful for comparison with the direct calculations of the yearly

83    [10]Be variations.  Newer [10]Be ice core measurements in Greenland (Berggren, et al., 2009) are

84    discussed in a separate paper (Webber, et al., 2010)

85    In the South Polar Region there are several [10]Be measurements of interest.  These include

86    measurements at Taylor dome (~78 degrees latitude) which have a basic time resolution ~2-3

87    years and cover the time period from ~1920 to 1990 (Steig, et al., 1996).  Also there is a recent

88    measurement by Moraal, et al., 2005, near SANAE (~70 degrees latitude) from a vertical face

89    covering the years from 1994 to 2003.  This latter measurement is of interest because the

90    analysis techniques appear to be similar to those used by Beer, et al., 1990, in their analysis of

91    the Dye 3 northern hemisphere measurements.   Also there is a measurement from an ice core

92    drilled at the South Pole in 1984 (Raisbeck, et al., 1990).  This measurement has a time

93    resolution ~8 years and is useful for confirming the average [10]Be concentration in the time period

94    from ~1939 to 1985.

## Comparing [10]Be Concentration Measurements and the Predicted Production

96    In Figure 1 we show the time sequence of the calculated yearly average values for [10]Be

97    production above 65 degrees latitude (little mixing) as well as the global average production

98    (extreme mixing) from 1939 to 2005 from the [10]Be production calculations of Webber, Higbie

99    and McCracken, 2007, using the individual year average values of the solar modulation

100    parameter as determined by Usoskin, et al., 2004, which is used to deduce the cosmic ray

101    spectrum at the top of the atmosphere (this calculation is discussed in detail in Webber, Higbie and McCracken, 2007).  Also shown in Figure 1 are the temporal measurements of [10]Be concentration in polar ice cores from Beer, et al., 1990 and Steig, et al., 1996.  The Beer data is in one year time periods as shown in Figure 2 of Beer, et al., 1990.  The Steig data is in 2-3 year intervals.   As noted in the introduction the statistical uncertainty of these individual measurements is ~5-10% for each data point.

A visual inspection of the Beer data shows an apparent 10-11 year periodicity which is in correlation with the predicted 11-year solar variation seen in Figure 1, perhaps delayed by ~1 year.  Note that the predicted 11-year production cycles do not have the same shape from cycle to cycle.  As shown in the production calculations in Figure 1, more sharply peaked cycles alternate with cycles with flat peaks, consistent with our understanding of the effects of the 22-year solar magnetic polarity changes in the overall solar modulation which affects the cosmic ray intensity.  These more subtle differences in the 11-year cycles are not evident in the [10]Be concentration data of Beer, et al., 1990.

For the Steig, 1996, data a simple visual inspection suggests a less than obvious correlation between measurements and predictions especially for solar cycle #19 starting in 1965 where the predicted [10]Be atmospheric production is a maximum in 1965 but the observations indicate a minimum [10]Be concentration.

## Comparing the Distributions of [10]Be Concentration Measurements and the Predicted Production

In Figure 2 we show the distributions of 1 or 2-3 year average absolute [10]Be concentration measurements.  At the bottom are the distributions from Steig, et al., 1996 in the southern hemisphere in blue (27x2-3 year intervals) and Beer, et al., 1990, 1998, in the northern hemisphere in red (47x1 year intervals).  In the center panel we show the distribution from Moraal, et al., 2005, in the southern hemisphere (10x1 year intervals) in black.  In the top panel the distribution of yearly average [10]Be production values on the polar plateau for the same 47 year time period, plus the additional later 10 year time period corresponding to the Moraal, et al., 2005, measurements is shown.   The scales of the [10]Be production and measured [10]Be concentration measurements are normalized so that an average polar cap production $=0.040/cm^2 \cdot s$ equals the average concentration $=1.20 \times 10^4$ atoms/g measured by Moraal, et al., (2005) for the 10 year period from 1994 to 2004.

132     Absolute differences in the northern hemisphere concentration measurements (average = 
133     $0.65 \times 10^4$ atoms/g for Beer, et al., 1990) and the southern hemisphere measurements, (average = 
134     $1.85 \times 10^4$ atoms/g for Steig, et al., 1996) and $1.20 \times 10^4$ atoms/g for Moraal, et al, 2005, are 
135     obvious in the lower two panels of Figure 2.  These differences of a factor ~2-3 have been 
136     discussed previously in the literature and attributed to differences in the characteristics of the 
137     precipitation in the N and S hemisphere (e.g., Beer, 2000).

138     What is unusual is the distribution of yearly average concentration levels measured at the 
139     Dye-3 location (Beer, et al., 1990).  This distribution of yearly averages shows an "extra" 
140     distribution of higher concentration levels (about 12 out of the total of 47 years) that is not part 
141     of the normal distribution, nor is it part of the yearly production distribution shown in the top 
142     panel for exactly the same time period.  This distribution of measured yearly averages for this 
143     time period is thus distinctly bi-model.

144     In Figure 3 we show a scatter plot of the yearly $^{10}$Be concentration measurements and the 
145     $^{10}$Be production.  This scatter plot shows more clearly the dichotomy between the "high" and 
146     "low" measurements and how they affect the slope of the regression curves obtained by 
147     comparing the Dye-3 yearly average ice core data with the production data.

148     Also we note that, in terms of the absolute $^{10}$Be concentration measurements, lower time 
149     resolution data from the South Pole averaged over the same 50 year time period from 1935-1985 
150     as the Steig, et al., 1996 measurements gives a $^{10}$Be concentration level = $3.35 \times 10^4$ atoms/g 
151     (Raisbeck, et al., 1990).  Lower time resolution data at Summit, Greenland in the northern 
152     hemisphere for roughly the same time period as the Beer, et al., 1990 measurements gives $^{10}$Be 
153     concentration levels ~$1.5 \times 10^4$ atoms/g (Yiou, et al., 1997). These N-S differences again illustrate 
154     the factor ~2-3 difference between N and S polar cap measurements over the same time period 
155     noted above.  If this is due to differences in the precipitation (snowfall) in the N-S hemisphere as 
156     suggested by Beer, 2000, then it represents a climatic effect that is several times larger than the 
157     typical average 11 year variation in $^{10}$Be production which is ~36% in the polar region.

**Discussion**

159     These long term differences between N and S polar $^{10}$Be concentrations are normally 
160     ascribed to different snow accumulation rates, R, between Greenland and Antarctica (e.g., Beer, 
161     2000).  The measured concentration C is then the ratio of the production (flux) F to R, C=F/R. 
162     Therefore the observed N-S polar differences are an example of a non-correlation of $^{10}$Be

163     production and [10]Be concentration, in this case due to climatic effects responsible for the
164     accumulation of [10]Be in the yearly ice core samples.

165        With regard to possible short term climatic or local effects on the "conversion factor"
166     between production and concentration, we especially wish to note the "high years" in the
167     distribution of concentrations measured by Beer, et al., 1990. Also in this same category near the
168     S pole are the "low years" of 1994 and 1998 as measured by Moraal, et al., 2005. These latter
169     authors conclude that these "low years" are due to "no known effect", but point out that there are
170     large differences in [10]Be concentration in two ramps taken only one year apart and differing in
171     location by only ~300m. We also note that in a study of short term changes in [10]Be
172     concentration during a single year, 2001, at Law Dome in Antarctica, Pedro, et al., 2006, find
173     that at least ~30% of the variance in the changes of the [10]Be concentration during this time is due
174     to "climatic (e.g., atmospheric) effects".

175        Returning to the "high years" measured by Beer, et al., 1990, it should be noted that, in
176     the high time resolution data for several years from a nearby site, Beer, et al., 1991, have found
177     many examples of [10]Be concentration variations that are uncorrelated with other variations, e.g.,
178     $\delta^{18}$O, on both short term, ~1 month, to longer term, 1-2 years. Thus in all of these cases we have
179     examples of the processes that determine the actual [10]Be concentration at any particular location
180     at any one time that are complex and vary on a short time scale and are not related to production
181     variations.

182        To make the above discussion more quantitative we have examined directly the cross
183     correlation between [10]Be production by cosmic rays and two of the [10]Be concentration yearly
184     measurements. In the first case, for the total yearly Beer, et al., 1990, 1998 data (47 years), the
185     cross correlation coefficient is 0.322 and the normalized regression line slope is 0.579.

186        In the second case, for the Steig, et al., 1996 data in 2-3 year intervals, the cross
187     correlation coefficient is 0.296 and the normalized regression line slope is 0.388.

188        Both of the cross correlation coefficients and the regression line slopes are unexpected
189     values for well correlated data sets. It is generally understood that values of simple cross
190     correlation coefficients between two data sets which are significantly less than ~0.5, such as the
191     ones we have calculated here, indicate a "poor" correlation (Hoel, 1954) and suggest therefore
192     that other factors in addition to the [10]Be production are important in the eventual [10]Be
193     concentration that is measured on both a short and longer time scale.

194    Also the regression line slopes ~0.4-0.6 between [10]Be concentration and [10]Be production
195    that we determine are much less than the slope of 1.0 that would be expected for a direct
196    proportionality between production and concentration.  This direct proportionality is frequently
197    assumed in historical projections of [10]Be concentration measurements and their relationship to
198    [10]Be production.

199    We note here also the work of Nikitin, et al., 2005, which finds the relationship between
200    cosmic ray measurements and [10]Be concentration is so weak that their conclusion is that
201    atmospheric processes play an essential role in violating the relationship between cosmic ray
202    fluxes and radionuclide concentration on both the long term and short term.

203    The above data along with many ongoing studies of climatic effects on [10]Be
204    concentration measurements are essential for understanding the [10]Be production-[10]Be
205    concentration relationship.  These climatic studies include those described by Pedro, et al., 2006,
206    as noted earlier and also Field, et al., 2006.  More recent studies by Heikkila, Beer and Feichter,
207    2008, and Field, Schmidt and Shindall, 2009, have evaluated these climatic effects in increasing
208    detail.  Field, Schmidt and Shindall have particularly noted that the modulation estimated by
209    McCracken, et al., 2004, for the Maunder minimum time period, based on a simple linear [10]Be
210    production-[10]Be concentration comparison over the last several hundred years, would be
211    significantly modified by climatic effects.

212    Using still another approach, comparing historical [14]C and [10]Be studies, Usoskin, et al.,
213    2009, have suggested that the solar (production) signal may dominate the [10]Be concentration
214    signal <u>only</u> on time scales longer than ~100 years while shorter term [10]Be concentration
215    measurements are greatly affected by local climate.

216    Due to the overall complexity of the relationship between [10]Be production and [10]Be
217    concentration measurements it is important to consider various different approaches in
218    evaluating the ability of the [10]Be ice core measurements to reproduce an accurate [10]Be
219    production record.

220    **<u>Summary and Conclusions</u>**

221    Recent new calculations of the expected production rates of [10]Be (Webber, Higbie and
222    McCracken, 2007) in the Earths atmosphere by cosmic rays based on measurements near the top
223    of the atmosphere on balloons and spacecraft over the last 60-70 years have enabled a direct
224    comparison to be made with polar ice core measurements of [10]Be concentrations covering essentially the same time period. This comparison reveals several levels of inconsistency between the calculated polar production and the yearly or 2-3 year [10]Be concentration measurements of Beer, et al., 1990 and Steig, et al., 1996, respectively, near the N and S poles as well as other ice core measurements in both the N and S hemispheres. These inconsistencies all suggest that "atmospheric" or "local" effects, which may be at least as large as or larger than changes in the input production function, are apparently able to modify the ratio of production to concentration of [10]Be. This result compliments recent climatic studies e.g., Field, Schmidt and Shindall, 2009, and references therein, which suggest climatic effects will produce significant modifications to the production to concentration ratio of [10]Be.

Specifically, we find in our study that (1); The cross correlation of the time series for individual 1-2 year intervals between [10]Be production and [10]Be concentration measurements in the time period 1939-1985 is quite low. In the case of N polar concentration measurements of Beer, et al., 1990, the cross correlation coefficient is ~0.322 and for S polar measurements of Stieg, et al., 1996, it is ~0.296. These low cross correlation coefficients as well as the large intrinsic scatter in the data points shown in Figure 3 suggest that factors other than the direct [10]Be production may be important in determining the actual [10]Be concentration that is measured for a particular year. The regression line slopes of between 0.4 and 0.6, (rather than the value of 1.0 for a 1:1 correspondence) from these same studies give a quantitative indication that the magnitudes of these other factors are comparable to the direct [10]Be production changes themselves.

(2); The data set for the N polar concentration measurements of Beer, et al., 1990, has a very unusual distribution with a second peak corresponding to yearly time intervals of "high" measured concentration, totally unlike the distribution of predicted yearly production for the same time period. The presence of these "high" concentration yearly intervals suggests that large changes apparently exist in the conversion factor between production and concentration measurements on time scales ~1-2 years or less.

(3); There is a persistent factor of 2-3 difference in N and S polar [10]Be concentration measurements over the years from ~1935-2005, with the S polar measurements being a factor ~2-3 times larger. Over the same 60-70 year time period the production calculations require that the same [10]Be production is experienced at both the N and S poles. These N-S differences have been noted and explained in terms of differences in snow accumulation rates (e.g., Beer, 2000).

256    Here we point out that, indeed if this is the sole explanation for these differences, then these
257    (atmospheric) effects are much larger than the $^{10}$Be production variations in an average 11-year
258    cycle.
259

**Figure Captions**

**Figure 1:** Calculated yearly average $^{10}$Be production above 65 degrees latitude for the years 1939-2005 (top panel); Global average production (2nd panel); yearly average $^{10}$Be concentration measurements above 65 degrees N latitude, Beer, et al., 1990, (3rd panel); 2-3 year average concentration measurements at 78 degrees S latitude, Steig, et al., 1996 (4th panel). The values for the production in the top two panels are x $10^{-2}$, for the bottom two panels the concentration values are x$10^4$ atoms/g, and 1.4 x $10^3$ atoms/g respectively.

**Figure 2:** Distributions of 1-2 year average $^{10}$Be concentration/production measurements. Bottom panel shows the concentration distribution from the Beer, et al., 1990, 1998 data (red) and Steig, et al., 1996 data (blue). Center panel is the distribution from the Moraal, et al., 2005, data. Top panel shows the distribution from the predicted polar $^{10}$Be production of Webber, Higbie and McCracken, 2007, normalized to the Moraal, et al., measurements 0.040/cm$^2$•s = 1.20x$10^4$ atoms/g (individual years 1994-2004 corresponding to the Moraal data are shown as a dotted line). Concentration distributions are off-set vertically by 16 (Moraal data) and 25 (predicted).

**Figure 3:** Scatter plot between the yearly average $^{10}$Be production above 65 degrees latitude and Beer, et al., 1990 $^{10}$Be ice core measurements for the "high" years (solid squares) and the "low" years (solid circles) as described in the text. The solid line is the best fit regression line for "all" data points with a slope of 0.579. The dashed line is for low points only, with a slope of 0.326.

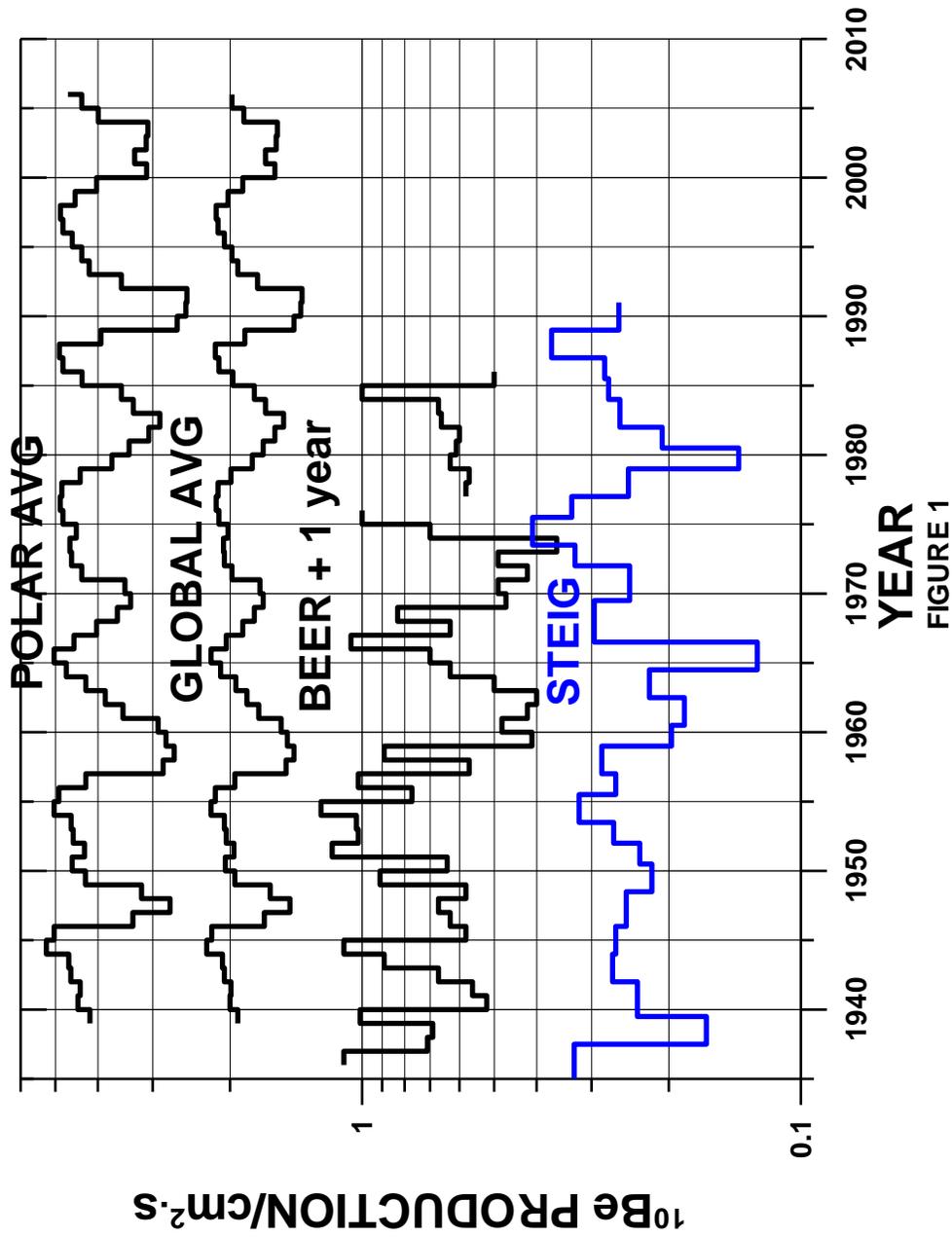

FIGURE 1

349

348

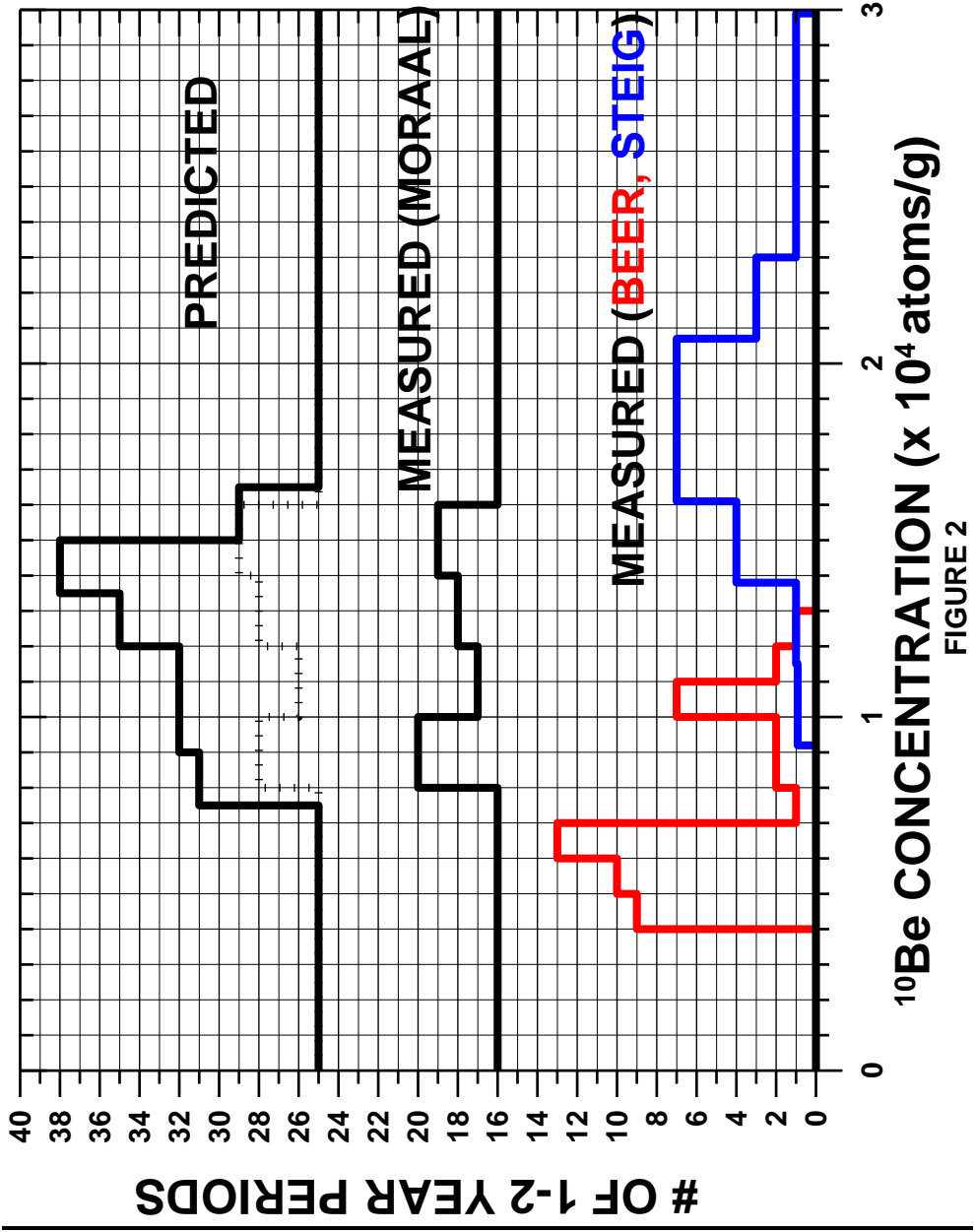

FIGURE 2

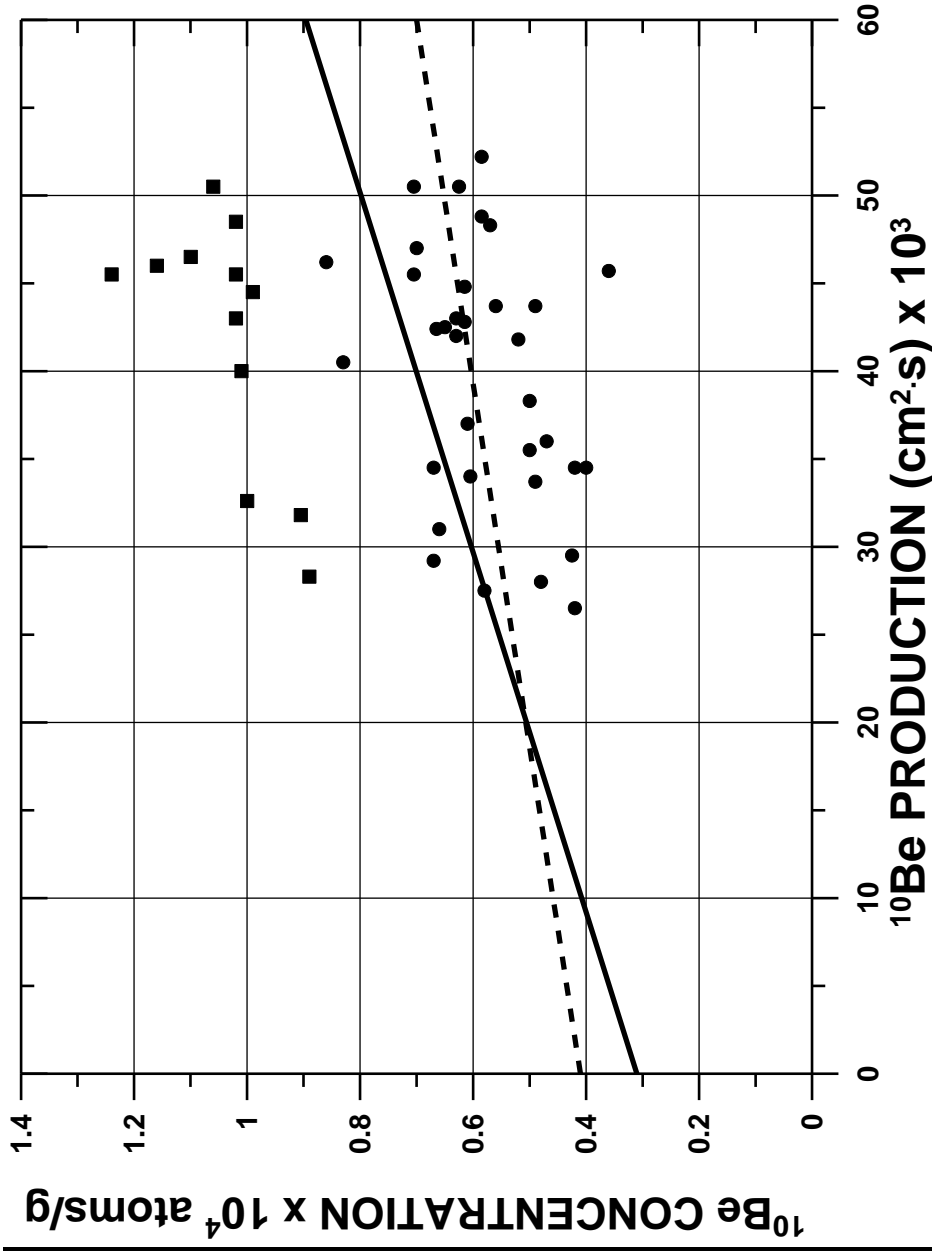

FIGURE 3